\documentstyle[preprint,aps]{revtex}
\input{epsf}
\begin{document}
\def\hatt{{\hat t}}
\def\hatx{{\hat x}}
\def\hatth{{\hat \theta}}
\def\hatta{{\hat \tau}}
\def\hatrh{{\hat \rho}}
\def\hatva{{\hat \varphi}}
\def\p{\partial}
\def\nn{\nonumber}
\def\tils{{\tilde s}}
\def\tila{{\tilde a}}
\def\bart{{\bar t}}
\def\barx{{\bar x}}
\def\barh{{\bar \rho}}
\def\np#1#2#3{Nucl. Phys. {\bf B#1} (#2) #3}
\def\pl#1#2#3{Phys. Lett. {\bf B#1} (#2) #3}
\def\prl#1#2#3{Phys. Rev. Lett.{\bf #1} (#2) #3}
\def\physrev#1#2#3{Phys. Rev. {\bf D#1} (#2) #3}
\def\ap#1#2#3{Ann. Phys. {\bf #1} (#2) #3}
\def\prep#1#2#3{Phys. Rep. {\bf #1} (#2) #3}
\def\rmp#1#2#3{Rev. Mod. Phys. {\bf #1}}
\def\cmp#1#2#3{Comm. Math. Phys. {\bf #1} (#2) #3}
\def\mpl#1#2#3{Mod. Phys. Lett. {\bf #1} (#2) #3}
\def\cqg#1#2#3{Class. Quantum Grav. {\bf #1} (#2) #3}
\preprint{hep-th/0105095}
\title{Super D-Helix}
\author{Jin-Ho Cho
\thanks{jhcho@taegeug.skku.ac.kr} and
Phillial Oh
\thanks{ploh@dirac.skku.ac.kr}}
\address{{\it BK21 Physics Research Division and Institute of Basic Science\\ Sungkyunkwan University, 
Suwon 440-746, Korea
}}
\date{\today}
\maketitle
\begin{abstract}
We study `Myers effect' for a bunch of $D1$-branes with $IIB$ superstrings moving in one direction along the branes. We show that the `blown-up' configuration is the helical $D1$-brane, which is self-supported from collapse by the axial momentum flow. The tilting angle of the helix is determined by the number of $D1$-branes. The radius of the helix is stabilized to a certain value depending on the number of $D1$-branes and the momentum carried by $IIB$ superstrings. This is actually T-dual version of the supertube recently found as the `blown-up' configuration of a bunch of $IIA$ superstrings carrying $D0$-brane charge. It is found that the helical $D1$ configuration preserves one quarter of the supersymmetry of $IIB$ vacuum.
\end{abstract}
\narrowtext
\section{Introduction}
$Dp$-branes interacting with higher form RR-fields ($(p+2n+1)$-form for example) become $D(p+2n)$-branes, which was first suggested by Emparan \cite{emparan} and nontrivial interactions were explicitly obtained by Myers \cite{myers}. In the presence of magnetic RR-fields, these interactions are caused by the motion of the branes. With the motion in compact space (thus carrying angular momentum), the higher dimensional brane increases its size. Since the size is bounded in the compact space, the angular momentum is also bounded. This fact possibly explains the stringy exclusion principle in the dual gravity setup \cite{susskind}. 

Recently Mateos and Townsend showed that this angular momentum can be given in a different guise \cite{townsend}. In some special setup of tubular $D2$-brane, the Poynting vector of the electromagnetic field on the $D2$ world-volume provides the angular momentum. This can be thought of as the `blown-up' configuration of a bunch of $IIA$ superstrings with $D0$-branes evenly distributed on it. It is self-supported from collapse by the angular momentum supplied by the electric and magnetic field associated with the number of $IIA$ superstrings and $D0$-branes respectively. The important fact is that the tube solution preserves one quarter of supersymmetry. This was generically possible for the intersecting $D$-branes with relative codimension four \cite{polchinski}.

In this paper we pursue the issue further to see how this effect can be understood in the T-dual setup. Several string duality transformations will yield straightforward generalizations of supertube. In Ref. \cite{townsend}, S-dual configurations (therefore of M theory) to the supertube were also discussed. These configurations will generate lots of lower dimensional descendants upon different compactifications, which are to be related with one another via U-dualities. Although these are the naive expectations, recent interests \cite{bak,emparan2,lunin,mateos} on this subject warrant to produce more explicit results. 

We first show that an array of $Dp$-brane $(p\leq 6)$ along some axis, say $X$-axis (see eq. (\ref{metric})), when threaded vertically by superstrings over the entire volume of $Dp$-branes, is `blown-up' to $D(p+2)$-brane of topology $R\!\!\!\!R^{p+1}\times S^1$ acquiring extra tubular two dimensions. This is obtained by taking T-duality along various directions transverse to $IIA$ superstrings carrying $D0$-brane charges. The radius of the circle $S^1$ is invariant under these T-dualities. An astonishing result is obtained when we take T-duality along the axial direction of the configuration. The bound state of $D0$-branes and $IIA$ superstrings becomes that of $D1$-branes with $IIB$ superstrings moving in one direction on it. We show that its corresponding `blown-up' configuration is a single helical $D1$-brane ($D$-helix) traveling with the speed of light along its axis. This is peculiar in that the dimensionality is not changed upon the `blowing-up'. This $D$-helix should be related with the helical $IIA$ string discussed in Ref. \cite{townsend} via a sequence of S- and T-duality.  

Before transferring to the next section, we start by briefly recapitulating the results of Ref. \cite{townsend}, where the configuration of $D0$-branes evenly arrayed along $X$-direction and threaded by a bunch of $IIA$ superstrings was considered. 

The configuration is embedded in flat geometry parametrized as
\begin{eqnarray}\label{metric}
ds^2=-dT^2+dX^2+R^2d\phi^2+dR^2+ds^2(E\!\!\!\!E^{(6)}).
\end{eqnarray}
Therefore it is free from any background gravitational effect and also there is no background field of any kind.
It was shown in Ref. \cite{townsend} that this can be considered as zero radius limit of tubular $D2$-brane. $D0$-brane charge is `dissolved' as magnetic flux on $D2$-brane while $IIA$ superstrings are dissolved as the electric field along $X$-direction. 

With the static gauge for the world-volume coordinates $(t=T,\, x=X,\, \varphi=\phi)$ on $D2$-brane, Born-Infeld (BI) 2-form field strength is given by 
\begin{eqnarray} 
F=E\,dt\wedge dx+B\,dx\wedge d\varphi.
\end{eqnarray}  
The Lagrangian for the tubular $D2$-brane is that of Dirac-Born-Infeld (DBI) simplified as
\begin{eqnarray} 
{\cal L}=-\sqrt{R^2(1-E^2)+B^2}.
\end{eqnarray}
For fixed momentum $\Pi\equiv\partial{\cal L}/\partial E$ and magnetic field $B$, the Hamiltonian ${\cal H}=R^{-1}\sqrt{(\Pi^2+R^2)(B^2+R^2)}$ is minimized at $R=\sqrt{|\Pi B|}$. The same physics can be viewed from $D0$-brane side. In this case, the system is described by the matrix model. Some extended solutions including multi-supertube configurations were found in Ref. \cite{bak}.

In the next section, we consider the cases of planar $Dp$-brane array threaded vertically by superstrings. These are obtained by T-dual transformation along various directions transverse to the $IIA$ superstrings carrying $D0$-brane charge. In section III, we study the case obtained by T-dual transformation along longitudinal direction of the original configuration. We show that the blown-up configuration is the helical $D1$-brane moving with light velocity along its axis. In section IV, we show explicitly $1/4$ of the supersymmetry of $IIB$ Minkowski vacuum is preserved in the $D$-helix configuration. In the last section we conclude with some discussions and remarks on further works.

\section{$Dp$-branes threaded by superstrings}

The first question that arises from the supertube physics is whether similar `blowing-up' happens in $IIB$ setup. One simple way to see this is to take T-duality along some directions for the original $IIA$ superstrings carrying $D0$-brane charge and check whether the same T-duality for the supertube gives sensible `blown-up' configuration. This is based on the fact that supersymmetry is preserved under T-duality \cite{alvarez}. Since the supertube configuration encodes those charges of $D0$-brane array and $IIA$ superstrings, so must do its T-dual counterpart. 

We first take T-duality along some directions transverse to $IIA$ superstring. The $D0$-brane array threaded by $IIA$ superstrings are dualized to be $D1$ array crossed by a bunch of $IIB$ superstrings. We expect $D3$-brane of topology $R\!\!\!\!R^2\times S^1$ as the `blown-up' configuration. This $D3$-brane is nothing but T-dual version of $IIA$ supertube. To be more specific, we take $X_4$-direction of $E\!\!\!\!E^{(6)}$ as the T-dual direction. 

Taking T-duality directly on DBI action is not simple. It is very obscure in the DBI action to start from $U(\infty)$ matrix and constrain its components (using orbifold technique used in Ref. \cite{douglas}) to describe $D$-brane array along some compact direction. Instead we take an indirect way. With the knowledge about BI fields on the $D3$-brane which encodes the dissolved $D1$-branes and $IIB$ superstrings, we construct DBI action for this $IIB$ setup. On the resulting $D3$-brane, the array of $D1$-branes is dissolved as magnetic flux and the number of $IIB$ superstrings is encoded as the electric field along $X$-direction. In the static gauge for the additional world-volume coordinate $x_4=X_4$, the flat geometry induced on the world-volume and the BI fields become
\begin{eqnarray} 
ds^2&=&-dt^2+dx^2+R^2d\varphi^2+dx_4^2,\nonumber\\
F&=&E\,dt\wedge dx+B\,dx\wedge d\varphi.
\end{eqnarray}

One can see $D3$ brane of topology $R\!\!\!\!R^2\times S^1$ is `blown-up' to have non-vanishing size of circle direction because DBI Lagrangian constructed from the above configuration is the same as that of the supertube. The basic reason why this gives the same results as $IIA$ case is that T-duality along $X_4$ preserves the relative codimension of $D2$-brane and $D0$-branes dissolved in it. Hence the same results will be obtained for further T-dualities along the directions along $X_{5,6,7,8,9}$. 

Summing up the result for an array of $Dp$-brane $(p\leq 6)$ along $X$-axis, we can say as follows; when threaded vertically by superstrings over the entire volume of $Dp$-brane, it is `blown-up' to $D(p+2)$-brane. The extra two dimensions acquired is tubular extended along $X$-direction and embedded in the residual dimensions. In all cases the stabilized radius is the same and governed by NS field $E$ of the superstrings, and the magnetic field $B$ which is produced effectively by $Dp$-branes dissolved in the $D(p+2)$-brane.

\section{$D1$-brane with traveling $IIB$ superstrings}

In this section we deal with another $IIB$ setup, by taking T-duality along $X$-direction. The basic question here is about the $IIB$ counterpart of the supertube. $D0$-brane array threaded by $IIA$ superstrings is T-dual to $D1$-branes along which $IIB$ superstrings are traveling in one direction. Since the former configuration is not stable in $IIA$ setup, neither should be the latter. At first sight, one might think this latter system will be blown up to $D3$-brane acquiring extra spherical two dimensions because there is no two dimensional object in $IIB$ setup. We show below that this is not the case. Actually the stabilized configuration remains one-dimensional. It turns out to be a $D$-helix with axial momentum flow, whose radius is the same as that of the supertube. 

The basic tool is again T-duality acting on the supertube. In order to see the resulting configuration, we study the boundary conditions of $IIB$ superstring. These can be obtained by T-dualizing the boundary conditions of $IIA$ superstrings living on the supertube;
\begin{eqnarray} 
IIA:\quad &&\left( \partial_\sigma X^0+E\partial_\tau X^1\right)|_{\sigma=0,\pi}=0,\nonumber\\
&&\left(\partial_\sigma X^1+E\partial_\tau X^0-B\partial_\tau\phi \right)|_{\sigma=0,\pi}=0,\nonumber\\
&&\left(R^2\partial_\sigma\phi+B\partial_\tau X^1 \right)|_{\sigma=0,\pi}=0.
\end{eqnarray}
T-duality along $X^1$-direction interchanges $\partial_\tau X^1$ with $\partial_\sigma X^1$. With $\tilde{X}^\mu$ denoting T-dualized coordinates, the above boundary conditions are T-dualized as
\begin{eqnarray} 
IIB:\quad &&\partial_\sigma\left(\tilde{X}^0+E\tilde{X}^1 \right)|_{\sigma=0,\pi}=0,\nonumber\\
&&\partial_\tau\left(\tilde{X}^1+E\tilde{X}^0-B\tilde{\phi} \right)|_{\sigma=0,\pi}=0,\nonumber\\
&&\partial_\sigma\left(R^2\tilde{\phi}+B\tilde{X}^1 \right)|_{\sigma=0,\pi}=0.
\end{eqnarray} 
From the second condition, we note the hypersurface $\tilde{X}^1+E\tilde{X}^0-B\tilde{\phi}=c$ (with the omitted conditions for other transverse directions) defines $D1$ world-sheet. We take the constant $c$ to be zero for simplicity. 

The other two conditions define the longitudinal directions of $D1$-brane. Since both of them are Neumann conditions, one can take arbitrary two independent combinations of the coordinates $\tilde{X}^0+E\tilde{X}^1$ and $R\tilde{\phi}+B\tilde{X}^1/R$ to make one temporal coordinate and one spatial coordinate. The simplest choice will be the `orthonormal' pair $(\tilde{X}^0,\,\, R\tilde{\phi})$. This choice is transparent if we see dual background geometry obtained by Buscher's duality \cite{buscher};

\begin{eqnarray} 
\tilde{ds^2}&=&-(d\tilde{X}^0)^2+(d\tilde{X}^1)^2+R^2(d\tilde{\phi})^2\nonumber\\
&=&-(d\bar{X}^0)^2+(d\bar{X}^1-E\,d\bar{X}^0+B\,d\bar{\phi})^2+R^2(d\bar{\phi})^2,
\end{eqnarray} 
where one can easily see the relation between the orthonormal coordinates $(\tilde{X}^0,\,R\tilde{\phi},\,\tilde{X}^1)$ and the tilted coordinates $(\bar{X}^0,\,R\bar{\phi},\,\bar{X}^1)$; 

\begin{eqnarray} 
\tilde{X}^0=\bar{X}^0,\qquad R\tilde{\phi}=R\bar{\phi},\qquad\tilde{X}^1=\bar{X}^1-E\bar{X}^0+B\bar{\phi}.
\end{eqnarray}

In the tilted coordinates, $D1$ brane is defined by the hypersurface $\bar{X}^1=0$. The other coordinates $(\tilde{X}^0,\, R\tilde{\phi})$ are orthonormal longitudinal coordinates. The metric induced on the hypersurface $(\bar{X}^1=0)$ becomes

\begin{eqnarray} 
\tilde{ds^2}=-\left(1-E^2 \right)(d\tilde{X}^0)^2-2EB\,\,d\tilde{X}^0d\tilde{\phi}+\left(R^2+B^2 \right)d\tilde{\phi}^2,
\end{eqnarray}
from which one gets DBI Lagrangian (in static gauge as $\hat{\tau}=\tilde{X}^0, \hat{\sigma}=\tilde{\phi}$);
\begin{eqnarray} 
{\cal L}=-\sqrt{ \left(1-E^2 \right)R^2+B^2}.
\end{eqnarray}

The DBI Lagrangian is of the same form as that of $IIA$ case. This is what we expected because T-duality in general leaves the $D$-brane action invariant. The only difference is the change of some field strength components into the derivatives of transverse scalar $\tilde{X}^1$, which now denote transverse fluctuations of $D1$-brane with respect to the world-sheet coordinates. In the case at hand, we can see this explicitly in the relation $\tilde{X}^1=-E\hat{\tau}+B\hat{\sigma}$.

\begin{eqnarray}
\frac{\partial \tilde{X}^1}{\partial\hat{\tau}}=-E,\quad \frac{\partial \tilde{X}^1}{\partial \hat{\sigma}}=B.
\end{eqnarray} 
Therefore $E$ is now the axial velocity of the $D$-helix. Fig.1 summarizes the configuration. 

As in the case of the supertube, the equations of motion just tells us that the momentum $\Pi$ is conserved. The momentum $\Pi$ stabilizes the radius of $D$-helix as $R=\sqrt{|\Pi B|}$, at which the velocity becomes the speed of light $E=\pm 1$ and Hamiltonian saturates its bound as ${\cal H}=|\Pi|+|B|$. One could consider the $IIB$ strings traveling along $D$-helix. However, uniform motion of those strings are not physical because of the world-sheet reparametrization symmetry. 

\epsfbox{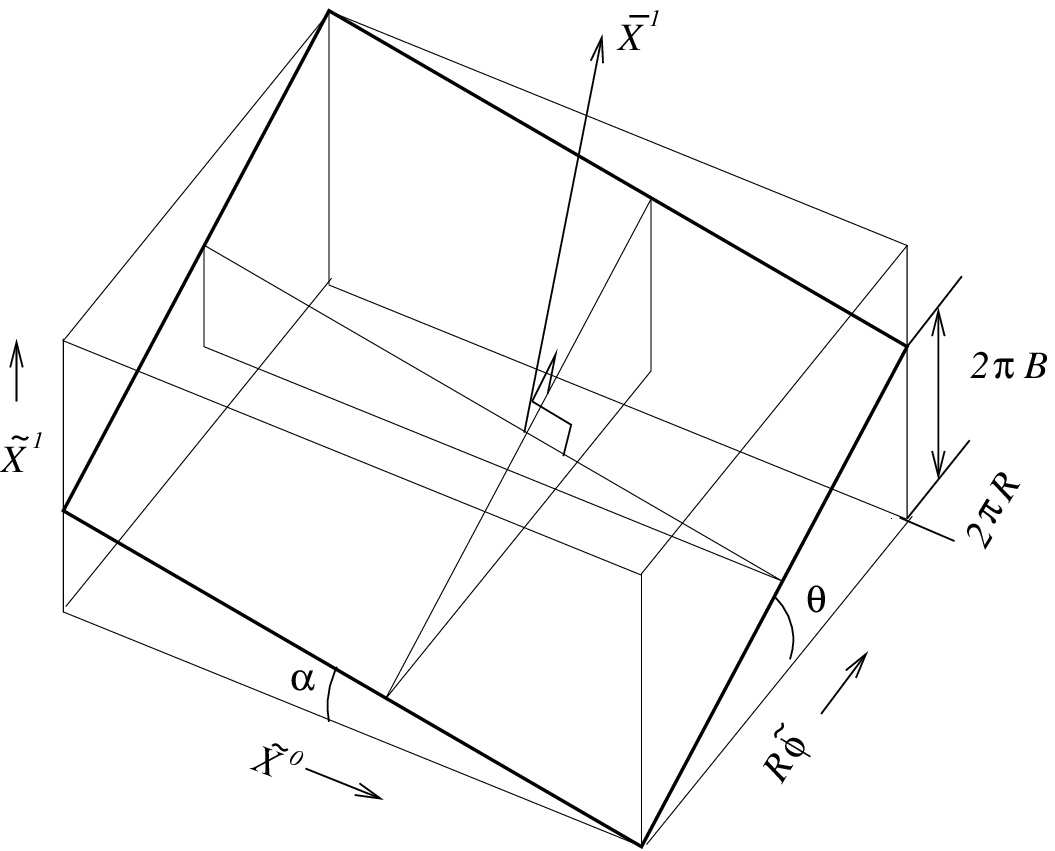}
\begin{flushleft}
{\footnotesize Fig. 1: The solid square represents $D1$ world-sheet. In the presence of $E=\tan{\alpha}$, $D1$-brane is not static. It is tilted with an angle of $\tan{\theta}=B/R$, thus helical. The helix pitch is $2\pi B$.}
\end{flushleft}

\section{Supersymmetry of the $D$-helix}
In this section we show that the above $D$-helix configuration preserves one quarter of supersymmetry as in the case of the supertube. This is, in fact, easy to understand because T-duality in general preserves supersymmetry \cite{alvarez}. However, it looks not so transparent to see supersymmetry directly in the $D$-helix configuration. Here we give a rigorous proof of that. We closely follow the procedure of Ref. \cite{townsend}. Supersymmetry is determined by the independent Killing spinors $\epsilon$ satisfying 
\begin{eqnarray}\label{killing} 
\Gamma\epsilon=\epsilon,
\end{eqnarray}
where $\Gamma$ is the matrix defining $\kappa$-transformation on the world-volume of $D$-branes \cite{bergshoeff}. In $IIB$ case at hand, it is 
\begin{eqnarray} 
\Delta\Gamma&=&\sigma_1\otimes\bar{\Gamma},\nonumber\\
\bar{\Gamma}&=&\left(B\Gamma_{\tilde{0}\tilde{1}} -E\Gamma_{\tilde{1}\tilde{\phi}}+\Gamma_{\tilde{0}\tilde{\phi}}\right),
\end{eqnarray}
where $\Delta\equiv\sqrt{-|\tilde{g}|}$ and static gauge is chosen.
Therefore the Killing spinor relation (\ref{killing}) becomes
\begin{eqnarray}\label{killing2} 
\left( 
\matrix{ 
0&\bar{\Gamma}\cr
\bar{\Gamma}&0\cr
}
\right)
\left( 
\matrix{
\epsilon^{1,\alpha}\cr
\epsilon^{2,\beta}\cr
}
\right)=\Delta
\left( 
\matrix{ 
\epsilon^{1,\alpha}\cr
\epsilon^{2,\beta}\cr
}
\right).
\end{eqnarray}

In $IIB$ case, chiral projection should be understood for the above spinors, so for $32$-component spinors $\epsilon^1$ and $\epsilon^2$. (See Ref. \cite{bergshoeff} for details.) Taking into account positive chirality components, we can put
\begin{eqnarray}\label{spinor} 
\epsilon^{1,\alpha}=\left( \matrix{\epsilon^{1,\alpha_1}\cr 0\cr }\right),\quad
\epsilon^{2,\alpha}=\left( \matrix{\epsilon^{2,\alpha_1}\cr 0\cr }\right).
\end{eqnarray}
Therefore they are effectively $16$-component spinors. 

In order to delimitate the coordinate dependent part of the spinors, we make use of the fact that Killing vectors can be written as bilinears of Killing spinors; $\xi^\mu=\bar{\epsilon}\,\Gamma^\mu\,\epsilon$ \cite{townsend2}. In flat geometry, Killing spinors can be written as
\begin{eqnarray}
\epsilon^{1,2}= e^{\frac{\tilde{\phi}}{2}\Gamma_{\tilde{R}\tilde{\phi}}}\,\,\epsilon_0^{1,2}, 
\end{eqnarray}
where $\epsilon_0$'s are constant spinors and $M_{\pm}=e^{\pm\tilde{\phi}\Gamma_{\tilde{R}\tilde{\phi}}/2}$ are Lorentz transformation acting on the chiral spinors. (Note that $\epsilon^{1,2}$ are of the same chirality.) With this in mind, one sees the Killing spinor relation (\ref{killing2}), i.e., $\bar{\Gamma}\epsilon^1=\Delta\epsilon^2$ becomes
\begin{eqnarray} 
M_+\left( B\Gamma_{\tilde{0}\tilde{1}}\epsilon^1_0-\Delta\epsilon^2_0\right)+M_-\left(\Gamma_{\tilde{0}\tilde{\phi}}-E\Gamma_{\tilde{1}\tilde{\phi}} \right)\epsilon^1_0=0.
\end{eqnarray}

In order to satisfy the above relation for arbitrary value of $\tilde{\phi}$, the first two terms and the last term should vanish separately;
\begin{eqnarray}\label{killing3} 
&&B\Gamma_{\tilde{0}\tilde{1}}\epsilon^1_0-\Delta\epsilon^2_0=0,\nonumber\\
&&\left(\Gamma_{\tilde{0}\tilde{\phi}}-E\Gamma_{\tilde{1}\tilde{\phi}} \right)\epsilon^1_0=0.
\end{eqnarray}
From the second relation, we see $\Gamma_{\tilde{0}\tilde{1}}\epsilon^1_0=-E\epsilon^1_0$, from which it is clear that $E=\pm 1$. (This was also obtained when we insert $R=\sqrt{|\Pi B|}$ into the expression of $E$.) Therefore $D$-helix should travel with the speed of light on its axis to make the configuration supersymmetric. With this value of $E$ inserted, the first relation of eq. (\ref{killing3}) tells us that
\begin{eqnarray} 
\epsilon^1_0=-sgn(B)\epsilon^2_0.
\end{eqnarray}
Hence $\epsilon^2_0$ can be written in terms of $\epsilon^1_0$. Since $\epsilon^1_0$ is constrained by the projection operator $\Gamma_{\tilde{0}\tilde{1}}$, only $8$ components of Killing spinors remain independent; thereby one quarter supersymmetry of $IIB$ vacuum are preserved for the configuration.

\section{Discussion}

In this paper, we studied the supertube physics in the T-dual setup. In the transversely T-dual case, the result looks somewhat plain; Blowing-up effect happens for an array of $D$-branes when they are threaded by superstrings. However, we have to mention one interesting point. In the case of $D1$ array crossed by superstring array, we can see why the stabilized radius is expressed by the product of $\Pi$ and $B$. If we take the S-duality, the role of $D1$-branes and that of superstrings are interchanged. In order for the radius to be invariant under S-duality, its square (by dimensional analysis) should be proportional to $\Pi B$. 

In the longitudinally T-dual case, we showed that a bunch of $D1$-branes with $IIB$ superstrings moving in one direction along the branes is `blown-up' to a $D$-helix traveling with the speed of light along its axis. It carries $1/4$ of the supersymmetry of $IIB$ vacuum.

One could consider $D3$-brane of topology $R\!\!\!\!\!R\times S^2$ as the blown-up configuration of the aforementioned $IIB$ case. However, that configuration is not consistent with $IIA$ result because it cannot be obtained by T-duality from the supertube configuration. In fact, when one construct DBI action by assuming flat geometry in the spherical polar coordinates and NS B field over the spherical part of the topology, the stabilized radius is vanishing. Spherically blown-up configurations can be possible only with appropriate background fields. Some examples are given in Ref. \cite{susskind} and other examples include $D$-branes embedded in the flux branes \cite{emparan2,cornalba}

The helical configuration of $D$-branes appeared in a different context. In Ref. \cite{hellerman}, a double helix connected to each other by fundamental strings is obtained by T-dualizing the $D$-brane setup of quantum Hall soliton \cite{susskind2}. With some additional input of $D$-branes (such as an array of $D6$-branes along $X$-axis, which is transverse to their world-volume, and T-duality as before) the super $D$-helix considered in this paper could be a starting point to study such a system further.

Finally, we mention two related problems to be solved. The first one concerns the well-known bound state of $D1$/$D5$-branes. As for an array of $D3$-branes threaded vertically by $IIB$ superstrings, one can apply the S-duality to change these superstrings into the same number of $D1$-branes. Subsequent T-dualities along the direction of $D1$-branes and another direction transverse to both branes result in the familiar $D1$/$D5$ bound state. The same sequence of dualities applied to the blown-up version of the initial configuration result in the Kaluza-Klein monopole-like configuration \cite{kaluza}. This is very puzzling. If blowing-up effect is preserved under U-duality, the result of supertube predicts that $D1$/$D5$ bound state is indeed unstable (even though it is known as a typical one-quarter supersymmetric state) and should be understood as a Kaluza-Klein monopole-like configuration. The same argument applies to the $D0$/$D4$ bound state, which is to be blown-up to $NS5$ brane wrapping a trivial homology cycle. This nontrivial observation cannot be easily understood by direct consideration of $D1$/$D5$ without recourse to S- and T-duality. Further study on this should be pursued.

The second one is another T-duality which is not considered in this paper. It is the T-duality along the angular direction of the supertube. This duality is quite intriguing as it involves a fixed point. Since the resulting geometry is singular, it should be taken with care. This will be studied elsewhere.

\section*{Acknowledgments}
We thank Soo-Jong Rey, Yoonbai Kim, and Youngjai Kiem for helpful discussions and Roberto Emparan for a valuable comment.
This work is supported in part by KOSEF through project NO. 2000-1-11200-000-3.


\begin{thebibliography}{99}
\bibitem{kaluza} It is of Kaluza-Klein monopole configuration (i.e., transversely T-dualized $NS5$-brane), but with one longitudinal direction compactifed in the external dimensions.
\end{thebibliography}
\end{document}